\documentclass[%
reprint,
superscriptaddress,
 amsmath,amssymb,
 aps,
 pra
]{revtex4-2}

\usepackage{graphicx}
\usepackage{dcolumn}
\usepackage{bm}
\usepackage{mathtools}
\usepackage{ulem}
\mathtoolsset{showonlyrefs,showmanualtags}
\usepackage{color}
\usepackage{comment}
\usepackage{xspace}
\usepackage{hyperref}

\newcommand{\X}{\mathcal{X}}

\newcommand{\Prob}{\mathbb{P}}

\newcommand{\Expect}{\mathbb{E}}

\newcommand{\rbra}[1]{\left(#1\right)}

\newcommand{\sbra}[1]{\left[#1\right]}

\newcommand{\reserve}[1]{}

\newcommand{\ecoli}{{\it E. coli }}
\newcommand{\Iup}{\mathcal{I}_{\mathrm{UB}}}
\newcommand{\Ilw}{\mathcal{I}_{\mathrm{LB}}}
\newcommand{\Iap}{\mathcal{I}_{\mathrm{TA}}}

\newcommand{\Signal}{\mathcal{X}}
\newcommand{\signal}{X}

\newcommand{\Dir}{E}

\newcommand{\M}{Y}

\newif\ifshowsections
\showsectionstrue  

\newcommand{\togglesections}{%
  \if\showsectionstrue\else\showsectionsfalse\fi
  \ifshowsections
    \let\section\oldsection
    \let\subsection\oldsubsection
  \else
    \renewcommand{\section}[1]{}
    \renewcommand{\subsection}[1]{}
  \fi
}

\newif\ifmaindoc
\newif\ifsuppdoc

\maindoctrue
\suppdocfalse

\newcommand{\SecEstimator}{S1}
\newcommand{\SecBinaryCase}{S2}
\newcommand{\SecBacteria}{S3}
\newcommand{\SecBootstrap}{S4}
\newcommand{\SecBinarization}{S5}
\newcommand{\SecFiniteSample}{S6}
\newcommand{\SecMotorTrajectory}{S7}
\newcommand{\FigInfoFitting}{S2}

\begin{document}

\preprint{APS/123-QED}

\title{
\ifmaindoc\else Supplementary Material for \protect\\\fi
Quantification of Information Flow by Dual Reporter System and Its Application to Bacterial Chemotaxis
}

\author{Kento Nakamura}
\altaffiliation{Correspondence: K. Nakamura (kento.nakamura@riken.jp; kento.nakamura.bio@gmail.com) \& T.J. Kobayashi (tetsuya@mail.crmind.net);
Also at Theoretical Sciences Visiting Program (TSVP), Okinawa Institute of Science and Technology Graduate University, Onna, 904-0495, Japan.}
\affiliation{%
RIKEN Center for Brain Science, 2-1 Hirosawa, Wako, Saitama 351-0198, Japan
}%
\author{Hajime Fukuoka}
\author{Akihiko Ishijima}
\affiliation{Graduate School of Frontier Biosciences, Osaka University, 1-3 Yamadaoka, Suita, Osaka 565-0871, Japan}
\author{Tetsuya J. Kobayashi}%
\altaffiliation{Correspondence: K. Nakamura (kento.nakamura@riken.jp; kento.nakamura.bio@gmail.com) \& T.J. Kobayashi (tetsuya@mail.crmind.net);
Also at Theoretical Sciences Visiting Program (TSVP), Okinawa Institute of Science and Technology Graduate University, Onna, 904-0495, Japan.}
\affiliation{Institute of Industrial Science, The University of Tokyo, 4-6-1, Komaba, Meguro-ku, Tokyo 153-8505 Japan}
\altaffiliation[Also at ]{Universal Biology Institute, The University of Tokyo, 7-3-1, Hongo, Bunkyo-ku, Tokyo 113-8654, Japan.}
\altaffiliation[Also at ]{%
Department of Mathematical Informatics, the Graduate School of Information Science and Technology, the University of Tokyo, 7-3-1, Hongo, Bunkyo-ku, Tokyo, 113-8654, Japan
}%




\date{\today}

\begin{abstract}
\ifmaindoc
Mutual information is a theoretically grounded metric for quantifying cellular signaling pathways.
However, its measurement demands characterization of both input and output distributions, limiting practical applications.
Here, we present alternative method that alleviates this requirement using dual reporter systems.
By extending extrinsic-intrinsic noise analysis, we derive a mutual information estimator that eliminates the need to measure input distribution.
We demonstrate our method by analyzing the bacterial chemotactic pathway, regarding multiple flagellar motors as natural dual reporters.
We show the biological relevance of the measured information flow by comparing it with theoretical bounds on sensory information.
This framework opens new possibilities for quantifying information flow in cellular signaling pathways.
\fi
\end{abstract}

\maketitle

\ifmaindoc

\paragraph*{Introduction}
Cells exhibit remarkable robustness in responding to changing environment despite the inherent stochasticity in sensing and signaling \cite{Rao2002-ru,Bowsher2014-ze,Francois2019-wk}.
Such robustness has led to the hypothesis that cellular signaling systems may operate at theoretically optimal levels under sensory and signaling noise \cite{Andrews2007-sy,Libby2007-zm,Tkacik2009-se,tostevin2009mutual,Petkova2019-zz,mora2019physical,Malaguti2021-rz,mattingly2021escherichia,Rode2024-ve,Tottori2024-dj,tkacik2025information}.
Supporting this idea, previous theoretical works have shown that biochemical reaction networks can implement information-theoretically optimal filtering \cite{kobayashi2010implementation,zechner2016molecular}, with the bacterial chemotactic network serving as an example of such possibility \cite{nakamura2021connection,nakamura2022optimal}.

We can test these optimality theories from the actual information flow measured by mutual information \cite{Shannon1948-zk,Bowsher2014-ze,Tkacik2016-dz,Uda2020-hw,Tang2022-vb,tkacik2025information}.
Mutual information is a theoretically grounded metric that connects sensing and signaling to decision-making and control, as exemplified by rate-distortion theory and stochastic thermodynamics \cite{Iglesias2016-nf,mattingly2021escherichia,Parrondo2015-is}.
For cellular systems, pioneering studies measured mutual information in the context of positional information during embryonic development of {\it Drosophila} \cite{Gregor2007-bd,Tkacik2008-bf}.
Building on these studies, researchers have quantified mutual information and related information metrics across a range of cellular systems, including bacterial chemotaxis \cite{mattingly2021escherichia}, yeast mating and stress responses \cite{Yu2008-qk,Granados2018-ye}, as well as mammalian and human immune signaling pathways, such as NF-$\kappa$B and ERK \cite{Cheong2011-gz,Uda2013-ux,Voliotis2014-ur,Selimkhanov2014-nv,Jetka2019-sp,Tang2021-vq,Achar2022-oa}.

Despite technical developments, measuring mutual information in actual biological systems remains challenging.
The difficulties arise primarily because mutual information demands characterization of the joint distribution of both input and output variables.
This challenge becomes especially pronounced when analyzing high-dimensional data, such as complex environmental signals or temporal trajectories \cite{Voliotis2014-ur,Selimkhanov2014-nv,Granados2018-ye,Jetka2019-sp,Tang2021-vq}.
Moreover, inferring the natural statistics of inputs is generally difficult, except in specific cases \cite{Gregor2007-bd,Tkacik2008-bf}.
While previous studies have circumvented this issue by assuming uniform input distributions \cite{Granados2018-ye} or by optimizing the input distribution to estimate channel capacity \cite{Cheong2011-gz,Uda2013-ux,Voliotis2014-ur,Selimkhanov2014-nv,Jetka2019-sp,Tang2021-vq,Achar2022-oa}, such assumptions may not represent the actual conditions under which cells operate.

To address these challenges, we propose an alternative method based on the dual reporter systems \cite{Swain2002-zw,Elowitz2002-vg,Hilfinger2011-dp,Bowsher2012-cq}.
Dual reporter systems enable identification of extrinsic fluctuations conveyed by the upstream pathway without specifying the extrinsic noise sources.
We extend this approach for information flow estimation that eliminates the need for measuring information sources or inputs.
We demonstrate the utility of our approach with the experimental data of the bacterial chemotactic pathway, where multiple flagellar motors function as natural dual reporters \cite{berg2004coli,tu2013quantitative}.

\paragraph*{Dual reporter system}
We formulate our approach using dual reporter systems, where the state of two reporter copies at time $t$, $Y_{t}$ and $Y'_{t}$, are driven by a common input signal $X$ (Fig.~\ref{fig:MIest_main}~(a)). In general, $X$ is high-dimensional and also $Y_{t}$ and $Y'_{t}$ are causally dependent on $\X_{t}:=X_{0:t}$, the past trajectory of $X$ up to $t$. 
The key physical properties underlying this system is that both reporters couple to the input identically while maintaining negligible direct interference between them.
This design allows us to assume two useful properties:
identical response statistics
\begin{align}
    \Prob_{Y_{t}\mid \X_{t}}(y\mid \X_{t}) = \Prob_{Y'_{t}\mid \X_{t}}(y'\mid \X_{t})\quad \mbox{for $y=y'$}, \label{eq:identity}
\end{align}
and conditional independence between reporters
\begin{align}
    \Prob_{Y_{t},Y'_{t}\mid \X_{t}}(y,y'\mid \X_{t})
    &= \Prob_{Y_{t}\mid \X_{t}}(y\mid \X_{t})\Prob_{Y'_{t}\mid \X_{t}}(y'\mid \X_{t}). \label{eq:conditional_independence}
\end{align}
Based on these properties and the law of total variation, we can estimate the extrinsic and intrinsic fluctuations originating from the input trajectory, defined by $\sigma_{\mathrm{ext}}^{2} := \mathbb{V}_{X_{0:t}}[\Expect_{Y_{t}\mid \X_{t}}[Y_{t}\mid \X_{t}]]$, and from the innate stochasticity of $Y_{t}$, $\sigma_{\mathrm{int}}^{2} := \Expect_{\X_{t}}[\mathbb{V}_{Y_{t}\mid \X_{t}}[Y_{t}\mid \X_{t}]]$.
These two fluctuations sum up to the total variation of $Y_{t}$: $\mathbb{V}_{Y_{t}}[Y_{t}]=\sigma_{\mathrm{ext}}^{2}+\sigma_{\mathrm{int}}^{2}$.
Notably, these quantities can be estimated without direct measurement of $\X_{t}$.
Indeed, for a set of reporter measurements $\{Y_{t_{i}},Y'_{t_{i}}\}_{i=1}^{N}$, the second moment of $\Expect_{Y_{t}\mid \X_{t}}[Y_{t}\mid \X_{t}]$ follows:
\begin{align}
    \mathbb{E}_{\X_{t}}\sbra{\Expect_{Y_{t}\mid \X_{t}}[Y_{t}\mid \X_{t}]^{2}}
    &= \mathbb{E}_{\X_{t}}\sbra{\Expect_{Y_{t}\mid \X_{t}}[Y_{t}\mid \X_{t}]\Expect_{Y'_{t}\mid \X_{t}}[Y'_{t}\mid \X_{t}]}\\
    &= \mathbb{E}_{\X_{t}}\sbra{\Expect_{Y_{t},Y'_{t}\mid \X_{t}}[Y_{t}Y'_{t}\mid \X_{t}]}\\
    &= \Expect_{Y_{t},Y'_{t}}[Y_{t}Y'_{t}]
    \approx \frac{1}{N}\sum_{i=1}^{N}Y_{t_{i}}Y'_{t_{i}}, \label{eq:second_moment_expectation}
\end{align}
where we sequentially apply Eq.~\eqref{eq:identity}, Eq.~\eqref{eq:conditional_independence} and the tower property of conditional expectation.
The final approximation holds for stationary and ergodic samples.

\paragraph*{Information flow estimation}

We extend this framework to estimate the mutual information between the input trajectory $\X_{t}$ and the instantaneous output $Y_{t}$:
\begin{align}
    \mathcal{I}[\X_{t};Y_{t}] = \Expect_{\X_{t}}\sbra{\sum_{y}\Prob_{Y_{t}\mid \X_{t}}(y\mid \X_{t})\log\frac{\Prob_{Y_{t}\mid \X_{t}}(y\mid \X_{t})}{\Prob_{Y_{t}}(y)}}. \label{eq:mutual_information}
\end{align}
Our strategy is to derive a quadratic approximation of the integrand $J := \Prob_{Y_{t}\mid \X_{t}}(y\mid \X_{t})\log [\Prob_{Y_{t}\mid \X_{t}}(y\mid \X_{t})/\Prob_{Y_{t}}(y)]$ with respect to $\Prob_{Y_{t}\mid \X_{t}}(y\mid \X_{t})$ and plug in estimators of its first and second moments.
A second moment estimator is derived by replacing $\Expect_{Y_{t}\mid \X_{t}}[Y\mid \X_{t}]$ in Eq.~\eqref{eq:second_moment_expectation} with $\Prob_{Y_{t}\mid \X_{t}}(y\mid \X_{t})$ (see Supplementary Mateiral (SM) Sec. \SecEstimator~for derivation):
\begin{align}
    \alpha_{2}(y)&:=\mathbb{E}_{\X_{t}}\sbra{\Prob_{Y_{t}\mid \X_{t}}(y\mid \X_{t})^{2}}\\
    &\approx \frac{1}{N}\sum_{i=1}^{N}\delta(Y_{i}-y)\delta(Y'_{i}-y)=:\hat{\alpha}_{2}(y). \label{eq:second_moment_probability}
\end{align}
The first moment estimator follows:
\begin{align}
    \alpha_{1}(y)&:=\mathbb{E}_{\X_{t}}\sbra{\Prob_{Y_{t}\mid \X_{t}}(y\mid \X_{t})}=\Prob_{Y_{t}}(y)\\
    &\approx \frac{1}{2N}\sum_{i=1}^{N}\{\delta(Y_{i}-y)+\delta(Y'_{i}-y)\}=:\hat{\alpha}_{1}(y). \label{eq:first_moment_probability}
\end{align}

We derive two complementary quadratic approximations of the mutual information (see SM Sec.~\SecEstimator~for derivation).
First, using a tangent of the $\log$ function, we obtain an upper bound:
\begin{align}
    \mathcal{I}[\X_{t};Y_{t}] \leq \Iup:=\sum_{y} \frac{\alpha_{2}(y)}{\alpha_{1}(y)} - 1. \label{eq:upper_bound_general}
\end{align}
Second, Taylor expansion around an arbitrary distribution $P_{0}$ yields
\begin{align}
     \mathcal{I}[\X_{t};Y_{t}] \approx \Iap := \mathcal{S}_{Y} - \mathcal{S}_{0} + \mathcal{D} + \frac{1}{2}\rbra{\sum_{y}\frac{\alpha_{2}(y)}{P_{0}(y)} - 1}, \label{eq:taylor_approx_general}
\end{align}
where $\mathcal{S}_{Y}:=-\sum_{y} \alpha_{1}(y)\log \alpha_{1}(y)$, $\mathcal{S}_{0}:=-\sum_{y} P_{0}(y)\log P_{0}(y)$, $\mathcal{D}:=\sum_{y} (\log P_{0}(y)+1)(\alpha_{1}(y)-P_{0}(y))$.
For binary state reporters $Y,Y'\in \{+1,-1\}$, this Taylor approximation $\Iap$ with uniform probability $P_{0}(y)=1/2$ coincides with a lower bound derived using the Pinsker inequality from the information theory (see SM Sec.~\SecBinaryCase~for detail):
\begin{align}
    \mathcal{I}[\X_{t};Y_{t}] \geq \Ilw = \Iap \label{eq:lower_bound_binary}
\end{align}
where $\Ilw := \mathcal{S}_{Y}-\mathcal{S}_{0}+2\{\alpha_{2}(y)-\alpha_{1}(y)\}+1/2\ (y\in\{\pm 1\})$.
Beyond the binary reporter case, we can generally use this result to bound from below the mutual information because binarization of reporter values decreases the mutual information due to information processing inequality.
We can now bound the mutual information from both above and below by substituting the estimators $\hat\alpha_{1}$ and $\hat\alpha_{2}$ given by Eqs.~\eqref{eq:second_moment_probability} and \eqref{eq:first_moment_probability} into Eqs.~\eqref{eq:upper_bound_general} and \eqref{eq:taylor_approx_general}, with reporter data binarized in advance when using Eq.~\eqref{eq:taylor_approx_general} as a lower bound.

\paragraph*{Application to bacterial chemotactic pathway}
\begin{figure*}[htbp]
    \centering
    \includegraphics[width=\linewidth]{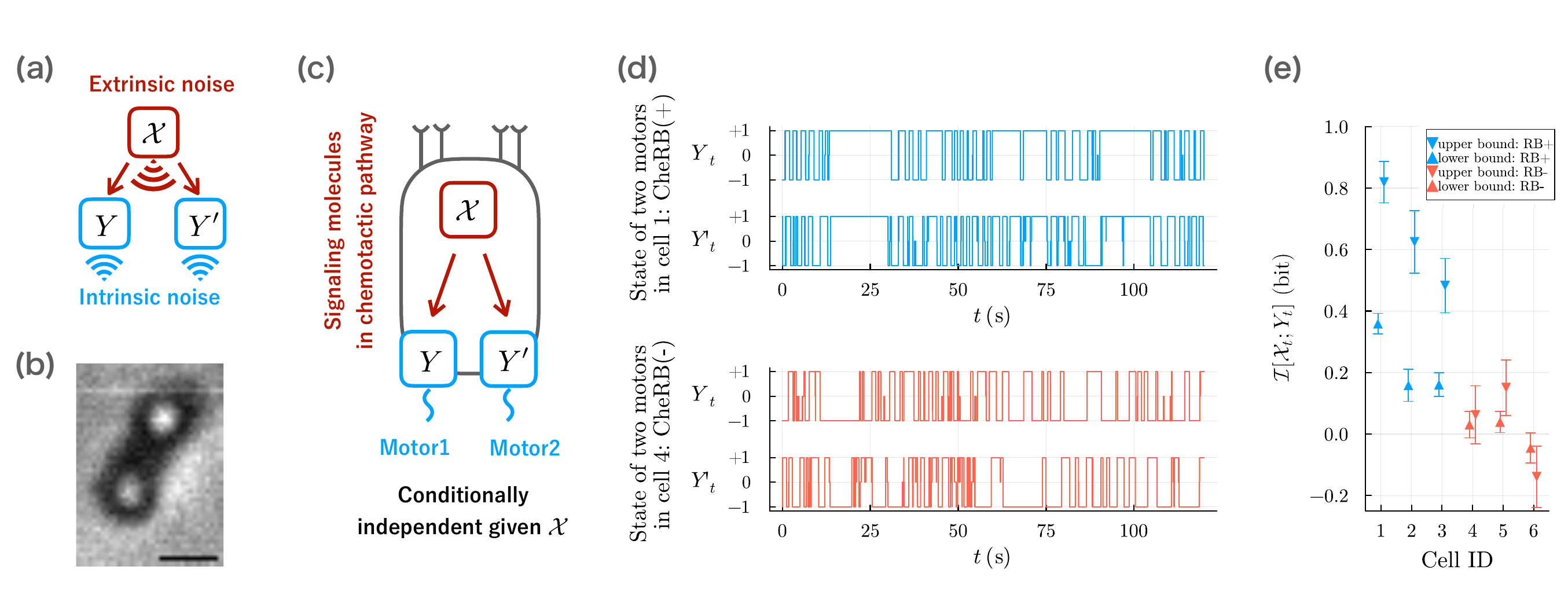}
    \caption{
    Dual reporter systems and mutual information estimation using bacterial flagellar motors as dual reporters.
    (a)
    Schematic of dual reporter systems. The total variance of reporters $Y$ decomposes into intrinsic noise $\sigma_{\mathrm{int}}^{2}$ from the innate stochasticity of $Y$ and extrinsic noise $\sigma_{\mathrm{ext}}^{2}$ originating from input signal $X$. Dual reporters enable estimation of extrinsic noise $\sigma_{\mathrm{ext}}^{2}$ without measuring input $X$.
    (b)
    Phase-contrast micrograph of an \ecoli cell with beads attached to two flagellar motors.
    Scale bar: $1$ $\mu \mathrm{m}$.
    (c)
    Schematic of two flagellar motors within a single cell functioning as dual reporters.
    Motor states are denoted by $\M_{t}$ and $\M'_{t}$, with input signal $\Signal_{t}$.
    (d)
    Representative time series of simultaneous measurements of two motors \cite{uchida2022chemoreceptor}, showing CCW ($+1$), stop ($0$), and CW ($-1$) states.
    Data shown for cell 1 (blue, functional chemotactic pathway, CheRB(+)) and cell 4 (red, impaired pathway, CheRB(-)).
    Complete data for all the cells are presented in SM Sec.~\SecMotorTrajectory.
    (e)
    Mutual information estimates for six cells: cells 1-3 (blue, CheRB(+), functional) and cells 4-6 (red, CheRB(-), impaired).
    Down- and up-pointing triangles show mean values of upper and lower bounds (Eqs.~\eqref{eq:upper_bound_general} and \eqref{eq:taylor_approx_general}) with error bars indicating $\pm 1$ standard deviation (SD) (see SM Sec.~\SecFiniteSample~for robustness to a choice of lower bound estimator).
    Statistics is computed using stationary bootstrap sampling \cite{politis1994stationary} ($10^3$ resamples, mean block length $10^{3.5}$; see Sec.~\SecBootstrap~for detail).
    Motor states were binarized by replacing stop states $\M_{t}=0$ independently for each motor and time point with $\M_{t}=\pm 1$ with equal probability (see Sec.~\SecBinarization~for robustness to a choice of binarization method).}
    \label{fig:MIest_main}
\end{figure*}


We demonstrate our approach in {\it Escherichia coli}'s chemotaxis signaling pathway.
\ecoli cells perform chemotaxis by sensing ligand gradients and regulating multiple flagellar motors.
The pathway accumulates the information of swimming direction, up or down the ligand gradient, obtained by sensory signal. Such information is stored and encoded in general into the trajectory of molecular states, $\signal_{0:t}$, in the pathway. The stored information in the trajectory should be decoded to efficiently control the current motor state $\M_{t}$ as the output.
We treat the multiple motors equipped with the cell as natural dual reporters to apply our method (Fig.~\ref{fig:MIest_main} (b,c)).
Previous studies established methods for simultaneous measurement of two motor copies, $\M_{t}$ and $\M'_{t}$, within a single cell \cite{Terasawa2011-zp,uchida2022chemoreceptor}.
As shown in Fig.~\ref{fig:MIest_main} (d), each motor alternates primarily between CCW (counter-clockwise; $+1$) and CW (clockwise; $-1$) states with the stop state (0) occurring infrequently.

Using this measurement data, we estimate the information flow from trajectory $\Signal_{t}:=\signal_{0:t}$ of upstream pathway molecules to the current motor state $\M_{t}$, quantified by $\mathcal{I}[\Signal_{t};\M_{t}]$.
To evaluate the lower bound based on Eqs. \eqref{eq:taylor_approx_general} and \eqref{eq:lower_bound_binary}, we binarize the motor states by replacing the infrequent stop state $\M_{t}=0$ with $\M_{t}= \pm 1$ with equal probability independently for each motor and time point.
To quantify the statistical uncertainty, we applied stationary bootstrap resampling \cite{politis1994stationary}, which preserves temporal correlations by sampling blocks of adjacent samples with lengths drawn from a geometric distribution (see SM Sec.~\SecBootstrap~for detail).

For cells with intact chemotactic signaling (CheRB+), we find lower and upper bounds ranging from $0.1$ to $0.4$ bit and $0.5$ to $0.8$ bit, respectively (Fig.~\ref{fig:MIest_main}~(e)).
These values align with the fact that a cell primarily requires the binary information of whether it is swimming up or down the gradient to control the motors.
However, these values may overestimate information transferred by the chemotactic signaling pathway, as upstream signal $\Signal_{t}$ could involve influence from other factors affecting both motors. Moreover, the values may reflect the interference between motors via the direct fluid dynamic intersection.
To address this ambiguity and verify that our estimates primarily reflect chemotactic signaling rather than confounding factors, we analyzed mutants with impaired chemotactic pathways (CheRB$-$; Fig.~\ref{fig:MIest_main}~(d)(bottom)).
As shown in Fig.~\ref{fig:MIest_main}~(e), the upper bound of mutual information in these mutants drops below $0.2\ \mathrm{bit}$, confirming that the higher information values observed in functional cells (CheRB+) predominantly result from chemotactic signaling.

\paragraph*{Sufficiency of the quantified information for chemotaxis}
We further evaluate the biological relevance of the estimated information $\mathcal{I}[\Signal_{t};\M_{t}]$ in the context of bacterial chemotaxis.
During chemotaxis, bacteria biases their random walk by regulating motors in response to sensed chemical gradients, which tells whether cells swim up or down the gradient.
This control mechanism requires information flow from environmental signals $\Dir_{t}$, representing swimming directions or chemical gradients, to motor states $\M_{t}$ via signaling pathway molecules $\Signal_{t}$.
We quantify this information flow using multivariate mutual information $\mathcal{I}[\Dir_{t};\Signal_{t};\M_{t}] = \mathcal{I}[\Dir_{t};\M_{t}] - \mathcal{I}[\Dir_{t};\M_{t}\mid \Signal_{t}]$ (Fig.~\ref{fig:sensory_information}~(a)).
From the chain rule of mutual information and its non-negativity, we derive the bounds $\mathcal{I}[\Dir_{t};\Signal_{t};\M_{t}]\leq \mathcal{I}[\Signal_{t};\M_{t}], \mathcal{I}[\Dir_{t};\Signal_{t}]$ (Fig. \ref{fig:sensory_information}~(b)).
These inequalities indicate that insufficient motor-side information flow $\mathcal{I}[\Signal_{t};\M_{t}]$ could limit the overall information flow and chemotactic performance, even if sensory-side information $\mathcal{I}[\Dir_{t};\Signal_{t}]$ is sufficient.
For optimal information transfer, the motor-side information should be comparable to the sensory-side information, minimizing information loss throughout the pathway (Fig.~\ref{fig:sensory_information}~(a)).

\begin{figure}[htbp]
    \centering
    \includegraphics[width=\linewidth]{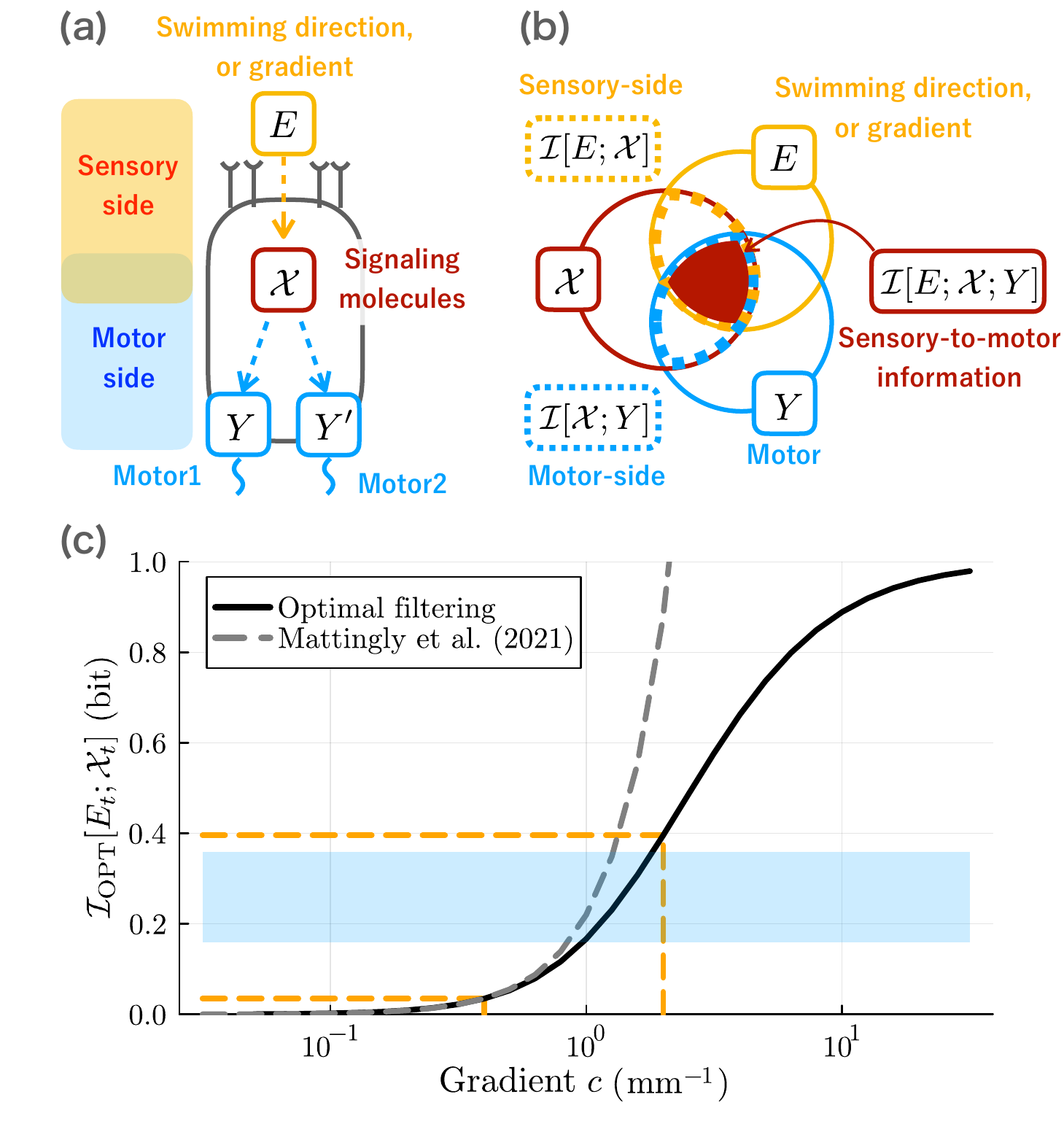}
    \caption{
    Information flow in bacterial chemotaxis.
    (a) Schematic of the chemotactic signaling pathway comprising swimming direction or gradient signals $\Dir_{t}$, signaling molecules $\Signal_{t}$, and two copies of motors $\M_{t},\M'_{t}$.
    (b) Diagram illustrating the inequality relationships between sensory-side information $\mathcal{I}[\Dir_{t};\Signal_{t}]$, motor-side information $\mathcal{I}[\Signal_{t};\M_{t}]$, and sensory-to-motor information $\mathcal{I}[\Dir_{t};\Signal_{t};\M_{t}]$.
    (c) Optimal sensory-side information $\mathcal{I}_{\mathrm{OPT}}[\Dir_{t};\Signal_{t}]$ as a function of gradient steepness $c$ with noise strength $\sigma$ fixed based on experimental information rate estimates.
    Gray dashed line shows experimental information rate estimates $\dot{I}_{s\rightarrow a}=\beta c^2$ ($\beta = 0.22$ bits s$^{-1}$~mm$^{2}$) \cite{mattingly2021escherichia}, shown after multiplying by the integration time $R^{-1} = 1$~s.
    Black solid line shows our filtering model prediction fitted to the weak gradient regime ($c = 0.1$-$0.4$ mm $^{-1}$) using noise strength $\sigma$ as a single free parameter (see Sec.~\SecBacteria~for detail).
    With physiological parameters $v=20\ \mu\mathrm{m}\ \mathrm{s}^{-1}$ and $R = 1\ \mathrm{s}^{-1}$ \cite{berg2004coli}, the fit yields $\sigma=6\cdot 10^{-4}$~s$^{-1}$.
    Dashed vertical orange lines indicate representative gradients $c=0.4$ and $2$~mm$^{-1}$ corresponding to weak and strong gradient regimes explored in previous experimental and theoretical studies \cite{mao2003sensitive,Englert2009-xn,jiang2010quantitative,flores2012signaling,mattingly2021escherichia,dufour2014limits}; horizontal dashed lines show the corresponding $\mathcal{I}_{\mathrm{OPT}}[\Dir_{t};Z_{t}]$ values.
    The shaded blue region indicates the range of lower bound for the motor-side information $\mathcal{I}[\Signal_{t};Y_{t}]$ in intact cells (ID 1-3) estimated in Fig.~\ref{fig:MIest_main}~(e).
    }
    \label{fig:sensory_information}
\end{figure}

To estimate the biologically relevant value of the sensory-side information, we draw on our previous work showing that bacterial sensory pathways likely implement Bayes filtering that maximizes mutual information by computing posterior probability \cite{nakamura2021connection,nakamura2022optimal}.
This result enables us to infer the upper bound of sensory-side information $\mathcal{I}_{\mathrm{OPT}}[\Dir_{t};\Signal_{t}]\geq\mathcal{I}[\Dir_{t};\Signal_{t}]$, which may be attained in actual sensory pathways.
At steady state, $\mathcal{I}_{\mathrm{OPT}}[\Dir_{t};\Signal_{t}]$ depends on a dimensionless signal-to-noise ratio (SNR) parameter $\lambda := 4c^{2}v^{2}/R\sigma$, which incorporates gradient steepness $c$, swimming speed $v$, tumble rate $R$, and noise strength $\sigma$.
While $v$, $R$, and $c$ are experimentally characterized, $\sigma$ has not been measured directly. 
We therefore fix $\sigma$ by requiring that the predicted information matches the recent experimental estimate of information rate $\dot{I}_{s\rightarrow a}$ in weak gradient regime ($c\leq 0.4$~mm$^{-1}$) \cite{mattingly2021escherichia}.
With this single fit, our filtering model reproduces the quadratic dependence of the information on $c$ in Ref. \cite{mattingly2021escherichia} at weak gradients and yields predictions for stronger gradients (Figs.~\ref{fig:sensory_information}(c) and \FigInfoFitting(a)).
Assumptions and derivations of these relationships are detailed in SM Sec.~\SecBacteria.

We compare sensory-side information across physiologically relevant gradient steepness $c$ with the lower bound of motor-side information estimated by our dual reporter method.
In the weak gradient regime ($c\leq 0.4$~mm$^{-1}$) \cite{mattingly2021escherichia}, sensory-side information ranges up to $0.04$ bits, value considerably below the motor information.
This suggests that motor side does not significantly limit information flow under weak gradient conditions.
Beyond the weak gradient regime, sensory-side information increases to approximately $0.4$ bits at steep gradient $c\sim 2$~mm$^{-1}$ explored in previous experimental and theoretical studies \cite{mao2003sensitive,Englert2009-xn,jiang2010quantitative,flores2012signaling,dufour2014limits}. This value only moderately exceeds the lower bound of motor information, suggesting that motor-side can be bottleneck in information transmission only at the upper range of physiologically relevant gradient steepness.

\paragraph*{Conclusion}
We introduced a new method for estimating mutual information that avoids direct input-output measurement by leveraging the dual reporter framework. Applying this method to bacterial flagellar motors as natural dual reporters, we quantified information flow in the chemotaxis pathway. Combined with theoretical estimates of sensory information, our results indicate that motor pathway may be tuned to match the maximum sensory information that cells encounter under physiological conditions.

A key limitation of our method lies in its strength: by not measuring the input, we cannot directly attribute the observed information to specific upstream sources. Nevertheless, as demonstrated here, targeted disruptions---such as pathway knockouts or source removal---allow estimation of their contribution. 
This is especially useful when experimental manipulation is feasible but direct quantification is difficult. In addition, if combined with conventional input-output methods, our method can reveal hidden information pathways: when our estimates of information exceed those from conventional ones, it may indicate the presence of additional, unmeasured routes of information flow.

This framework can be extended in several directions. One is the estimation of other information-theoretic quantities, such as transfer entropy~\cite{mattingly2021escherichia,Mattingly2024-nf}. Another is refinement via higher-order corrections using multiple reporters. With $k$ reporter copies $Y_t^{(1)}, \ldots, Y_t^{(k)}$ satisfying the dual reporter assumptions, we can estimate $\alpha_{k}(y):=\Expect_{\X_{t}}[\Prob_{Y_{t}\mid \X_{t}}(y\mid \X_{t})^{k}]$, which enables higher-order corrections to our approximation (see SM Sec.~\SecEstimator). As single-molecule techniques advance, opportunities to observe multiple intracellular copies increase, allowing such multi-reporter approaches. 

Our approach and these extension along with the conventional methodology will expand the analytical toolkit for advancing our understanding of cellular information processing across diverse biological systems.

\paragraph*{Acknowledgements}
We thank Takehiro Tottori for fruitful discussion.
The first author received a JSPS Research Fellowship (Grant Number 20J21362). This research was supported by JSPS KAKENHI (Grant Number 19H05799 and 25H01365) and JST CREST (Grant Number JPMJCR2011).This research was partially conducted while visiting the Okinawa Institute of Science and Technology (OIST) through the Theoretical Sciences Visiting Program (TSVP).


\else\fi
\bibliographystyle{apsrev4-2}
\providecommand{\noopsort}[1]{}\providecommand{\singleletter}[1]{#1}

\end{document}